\def\year{2020}\relax
\documentclass[letterpaper]{article} 

\usepackage{aaai20}  
\usepackage{times}  
\usepackage{helvet}  
\usepackage{courier}  
\usepackage{array}
\usepackage{mathtools}
\usepackage{multirow}
\usepackage{enumitem}
\usepackage{dsfont}
\usepackage{url}  
\usepackage{graphicx, subfigure}
\usepackage{amsmath}
\usepackage{amsfonts}
\usepackage{bm}
\usepackage{amssymb}  
\usepackage[draft]{todonotes}

\usepackage[linesnumbered,boxed,ruled]{algorithm2e}

\usepackage{amsmath} 
\usepackage{ stmaryrd }
\usepackage{diagbox}  

\newcommand{\nosection}[1]{\vspace{2pt}\noindent\textbf{#1.}}
\begin{document}
\title{Privacy Preserving PCA for Multiparty Modeling}

\author{
Yingting Liu\textsuperscript{1,2} , Chaochao Chen\textsuperscript{2}, Longfei Zheng\textsuperscript{2}, Li Wang\textsuperscript{2}, Jun Zhou\textsuperscript{2}, Guiquan Liu\textsuperscript{1}\thanks{Guiquan Liu is the corresponding author.}, Shuang Yang\textsuperscript{2}\\
~\textsuperscript{1}University of Science and Technology of China\\
~\textsuperscript{2}Ant Financial Services Group\\
~sa517218@mail.ustc.edu.cn, gqliu@ustc.edu.cn\\
~\{chaochao.ccc, zlf206411, raymond.wang, jun.zhoujun, shuang.yang\}@antfin.com
}

\maketitle

\begin{abstract}
    In this paper, we present a general multiparty modeling paradigm with Privacy Preserving Principal Component Analysis (PPPCA) for horizontally partitioned data. 
    PPPCA can accomplish multi-party cooperative execution of PCA under the premise of keeping plaintext data locally. We also propose implementations using two techniques, i.e., homomorphic encryption and secret sharing.
    The output of PPPCA can be sent directly to data consumer to build any machine learning models. We conduct experiments on three UCI benchmark datasets and a real-world fraud detection dataset. Results show that the accuracy of the model built upon PPPCA is the same as the model with PCA that is built based on centralized plaintext data.
\end{abstract}

\section{Introduction}

    Although big data concepts and technologies have become popular, a single organization is likely to have limited data, which makes it difficult to meet the requirements of machine learning models. Meanwhile, different organizations are often unwilling or unable to share data because of the competition or privacy regulations. \textit {Isolated data islands} situation has become a serious problem weakening the performance of artificial intelligence in the real world.
    
    Existing researches adopt cryptographic techniques, e.g., homomorphic encryption \cite{hardy2017private} or secure Multi-Party Computation (MPC) \cite{mohassel2017secureml}, to solve these problems. However, they have two weaknesses. First, their running time is much longer than the models built on aggregated data due to high computation complexity and communication complexity of the cryptographic techniques. It gets even worse in practice because of the large scale data. Second, they are usually model-specific, i.e., the models need to be built case by case. However, in real-world applications such as fraud detection and risk control, machine learning models usually need to be rebuilt periodically. If a general modeling or preprocessing method can be designed, the deployment of privacy preserving machine learning models can be greatly accelerated and the development costs can be significantly reduced. 
    
    
    Furthermore, Principal Component Analysis (PCA) is popularly used to transform the original data from one space to another \cite{pearson1901liii}. 
    Under isolated data island situation, how to perform PCA becomes a hot research problem. For example, Al-Rubaie \textit{et al} \cite{al2017privacy} proposed to use additive homomorphic encryption and garbled circuits for PCA with a single data user. With the increase in feature dimension, the computing time increases rapidly. Grammenos \textit{et al} \cite{grammenos2019federated} adaptively estimated the rank of PCA with unknown distribution in a streaming setting. Liu \textit{et al} \cite{liu2018privacy} proposed a huge digital data outsourcing scheme. However, none of them focus on building privacy preserving PCA for multi-party modeling.

    
    For these reasons, we propose a general modeling paradigm with PPPCA for horizontally partitioned data, which belongs to the category of \textit{shared machine learning} \cite{chen2018privacy,chen2020practical}. It involves two methods using homomorphic encryption and secret sharing respectively.
    Results of PCA can protect data privacy to a certain extent, and can be further improved by using differential privacy \cite{wei2016analysis}. Under the honest-but-curious setting, the results of PPPCA can be sent directly to any data consumer to build any machine learning models. Our contributions are two-folds: 
    
    \begin{itemize}
    
        \item
        We propose a privacy preserving modeling framework using PPPCA, where the plaintext data are hold by participants locally. The result of PPPCA can be carried out for building any  machine learning models, which significantly improves the generality and efficiency compared with the models built using time-consuming cryptographical techniques case by case.
    
        \item
        We propose two PPPCA approaches using secret sharing and homomorphic encryption respectively. Experiments on three UCI benchmark datasets and a real-world fraud detection dataset demonstrate that our approaches are lossless comparing with the approach that aggregates plaintext data.
         
    \end{itemize}
    

\section{Preliminaries}
    
    In this section, we present problem statement and some preliminary concepts of our proposal, including secret sharing and homomorphic encryption.
   
    \subsection{Problem Statement}
            
            \nosection{Data Setting} 
            Data provided by different participants are partitioned horizontally in the data space. That is, datasets share the same feature space but different sample space.
            
            
            \nosection{Input} 
            The plaintext data that are provided by $M$ $(M>1, M \in Z^+)$ participants.
            We use $\boldsymbol{X_i}$ ($1 \leq i \leq M$) to denote the corresponding plaintext data of $i$-th participant, where each column represents an attribute, and each raw represents a sample. In horizontally partitioned situation, the joint data can be formalized as $\boldsymbol{X} = \left (\begin{smallmatrix} \boldsymbol{X_1}^T & \boldsymbol{X_2}^T & \cdots & \boldsymbol{X_M}^T \end{smallmatrix}\right)^T $. 
            We use $d_i$ and $n_i$ to denote the feature dimension and the sample dimension of participant $i$, and use $d$ and $n$ to denote the feature dimension and sample dimension of the joint data, where $n=\sum^M_{i=1} n_i,d=d_1=\cdots=d_M$. 
            
            \nosection{Output} 
            The dimension reduced matrix $X'_i$ from each participant whose dimension is $n_i \times k$ $(k<d) $. It can be carried out for building any machine learning models.
            
            \nosection{Constraint} The lossless \& privacy preserving ability. The performance of PPPCA must be comparable to the non-private solution that brings all data in one place, while protecting the data privacy.

    \subsection{Additive Homomorphic Encryption}
        Additive homomorphic encryption, e.g., paillier \cite{paillier1999public}, is a method that supports secure addition of numbers given only the encryptions of them. The ciphertext operation on untrusted participant is secure. The result of the operation needs to be sent to the private key holder for decryption to get the plaintext result. We denote the ciphertext of a number $u$ as $\llbracket u \rrbracket$ and overload `$+$' as a homomorphic addition operation on ciphertext space. Additive homomorphic encryption satisfy that for any plaintext $u$ and $v$, 
            \begin{equation}
            \llbracket u \rrbracket + \llbracket v \rrbracket = \llbracket u+v \rrbracket.
            \end{equation}
        However, plaintext in this arithmetic must be the element of finite field. In order to support floating point operations, we adopt the encoding scheme \cite{hardy2017private}. This approach is based on a floating-point representation, and ultimately encode a number as a pair consisting of an encrypted significant and an unencrypted exponent. Moreover, these operations can be extended to work with vectors and matrices component-wise.

    \subsection{Secret Sharing}
        Secret sharing is a technique of secure Multi-Party Computation (MPC), which is widely used to build machine learning models \cite{mohassel2017secureml,chen2020secure}. In our case, all data providers participate in the calculation together, and the final result is reconstructed from their calculation results. Therefore, we choose n-out-of-n secret sharing \cite{singh2017secure}, which is an efficient scheme. 
        
        In this paper, we mainly focus on additive secret sharing. To additively share $l$-bit secret value $s$ owned by party $M$, party $M$ first generates $M$ shares $\{r_1, \cdots, r_{M}\}$, where $ r_i \in \mathds{Z}_{2^l} $ ($1 \leq i \leq M-1 $) are random numbers generated using Pseudo-Random Generator (PRG) and the last share $r_M=s- \sum_{i=1}^{M-1}r_i$ mod $2^l$. Then party $M$ distributes these shares to $M$ data providers (including itself). All intermediate values are secret-shared between data providers. We denote the share owned by party $i$ as ${\langle s \rangle}_i$. Given secret values $\{s_1, \cdots, s_M\}$ from $M$ data providers, to calculate the sum of them, each data provider needs to calculate the sum of their local shares (e.g., party $i$ calculates intermediate values $\langle v_i \rangle = \sum_{j=1}^M {\langle s_j \rangle}_i $ mod $2^l$). Finally, the result is the sum of these local results. During this process, all participants are invisible to others' inputs.
        
        The above secret sharing schema works in finite field, and can be extended to real number field following the encode and denote scheme in \cite{hardy2017private}. It is trivial to generalize the single number operations to matrices.
    

\section{The Proposed Method}

    In this section, we propose PPPCA framework and its two implementations. 
    
    \subsection{Overview}
        PPPCA aims to perform private PCA for multiple data providers under server-aided setting.
        It refers to three roles: data provider, server, and data consumer. Data provider owns the data and participates in the calculation. Server does not have any input, participates in some computation but does not receive any privacy output. It can be untrusted but cannot collude with data providers. Data consumer use the result of PPPCA for further modeling. 
        It takes three steps to execute PPPCA under horizontally partitioned case:
        
        \begin{itemize}
    
            \item Firstly, all data providers jointly normalize the columns (attributes) to zero with secure operation. For example, each value of attribute $t$ owned by different data providers should be divided by the mean of them, which is denoted as $\overline{x_t}=\frac{\sum_{i=1}^M s_{it}}{n}=\frac{\sum_{i=1}^M \sum_{j\in I_i}x_{jt}}{\sum^M_{i=1} n_i}$, where $s_{it}$ is the sum of observations of feature $t$ owned by data provider $i$ and can be computed locally. $\overline{x_t}$ can be computed by all participants using secure additive operation.
            
            \item Secondly, all data providers jointly compute the the covariance matrix $\boldsymbol{C}$:
                \begin{equation}\begin{split}
                \boldsymbol{C}&=\frac{1}{n-1}\boldsymbol{X}^T\boldsymbol{X}\\
                &=\frac{1}{n-1}\left (\begin{matrix} \boldsymbol{X_1}^T & \boldsymbol{X_2}^T & \cdots & \boldsymbol{X_M}^T \end{matrix}\right)  \left (\begin{matrix} \boldsymbol{X_1} \\ \boldsymbol{X_2} \\ \vdots \\ \boldsymbol{X_M} \end{matrix}\right)\\
                &=\frac{1}{n-1} (\boldsymbol{X_1^TX_1} + \boldsymbol{X_2^TX_2} + \cdots + \boldsymbol{X_M^TX_M}), 
                \end{split}\end{equation}
            where $\boldsymbol{X_i}^2=\boldsymbol{X_i}^T\boldsymbol{X_i}$ can be computed locally. Hence, $\boldsymbol{C}$ can be computed by all data providers and server collaboratively using secure additive operation. At the end of this step, the plaintext of $\boldsymbol{C}$ is owned by server.
    
            \item Thirdly, server calculates the eigenvalues and corresponding eigenvectors of $\boldsymbol{C}$ locally, chooses $k$ $(k<d) $ eigenvectors to form the matrix $\boldsymbol{T}$ according to the size of the eigenvalues, and broadcasts it to data providers. Each data provider calculates $\boldsymbol{X'_i}=\boldsymbol{X_iT}$ locally and sends it to data consumer for further modeling.
        
        \end{itemize}
        
        Step 1 and step 2 will use secure operation, which can be boiled down to secure matrix addition operations. We will present how to implement these two steps with homomorphic encryption and secret sharing. 
        
    \subsection{Additive Homomorphic Encryption Based Method}
        Secure additive operation is done by Homomorphic Encryption (HE). The server is responsible for generating a key pair, sharing public key with all participants, and holding the private key for decryption. For secure matrix additions in step 1 and 2, all data providers firstly encrypt matrices that need to be added (e.g., $\boldsymbol{X_i}^2$) and send them to one of the data provider (denoted as $p$, $1 \leq p \leq M-1 $). Party $p$ receives these encrypted matrices and does the homomorphic additive operation in ciphertext space. Finally, party $p$ sends the result to server for decryption. 
        We present the whole HE based PPPCA framework in Algorithm \ref{alg: he-part-p}.
        
            
            
            
        
        \begin{algorithm}[t]
        	\LinesNumbered 
            \SetKwInOut{Input}{Input}\SetKwInOut{Output}{Output}
        	\Input{Data $\{\boldsymbol{{X}_{1}}, \cdots, \boldsymbol{{X}_{M}}\}$ ; dimension after PCA ($k$)}
        	\Output{Dimension reduced matrix $\boldsymbol{X'}$}
            
            Server generates a key pair and sends public key to all data providers;\\
            
            Data providers compute and encrypt the mean of their features locally as $\llbracket \mathbf{S_{i}} \rrbracket$ and send it to party $p$;\\
            
            Party $p$ receives $\{ \llbracket \mathbf{S_{1}}\rrbracket, \cdots, \llbracket \mathbf{S_{M}} \rrbracket \}$, computes $\llbracket \overline{x} \rrbracket = \sum_{i=1}^M \llbracket \mathbf{S_{i}} \rrbracket$, and sends it to server;\\
            
            Server receives $\llbracket \overline{x} \rrbracket$, decrypts it, and sends it to all data providers;\\
            
            Data providers receive $\overline{x}$ and normalize columns to zero locally;\\
            
            Data providers compute and encrypt $\llbracket \boldsymbol{C_i} \rrbracket = \llbracket \frac{1}{n-1}\boldsymbol{X_i}^T\boldsymbol{X_i} \rrbracket$ locally and send it to party $p$;\\
            
            Party $p$ receives $\{ \llbracket \mathbf{C_{1}}\rrbracket, \cdots, \llbracket \mathbf{C_{M}} \rrbracket \}$ , computes $\llbracket \boldsymbol{C} \rrbracket = \sum_{i=1}^M \llbracket \mathbf{C_{i}} \rrbracket$, and sends it to server;\\
            
            Server decrypts $\llbracket \boldsymbol{C} \rrbracket$, computes transfer matrix $\boldsymbol{T}$, then sends it to all data providers;\\
            
            Data providers receive matrix $\boldsymbol{T}$, calculate $\boldsymbol{X'_i} = \boldsymbol{X_iT}$ locally, then send it to data consumer;\\ 
            
            Data consumer receives dimension reduced matrix $\boldsymbol{X'}={(\boldsymbol{{X'_{1}}^T}, \cdots, \boldsymbol{{X'_{M}}}^T)}^T$ for modeling;\\
            
        	\Return{Dimension reduced matrix $\boldsymbol{X'}$};
        	\caption{HE based PPPCA framework}
        	\label{alg: he-part-p}
        \end{algorithm}

    \subsection{Secret Sharing Based Method}
        In this method, secure additive operation is done by secret sharing. In the last step of secret sharing, all of the data providers will send their local calculation results on shares to server.
        Then server computes the sum of these shares to reconstruct the plaintext result. 
        We present the whole Secret Sharing (SS) based PPPCA framework in Algorithm \ref{alg: ss-part-p}.

        
        \begin{algorithm}[t]
        	\LinesNumbered 
            \SetKwInOut{Input}{Input}\SetKwInOut{Output}{Output}
        	\Input{Data $\{\boldsymbol{{X}_{1}}, \cdots, \boldsymbol{{X}_{M}}\}$ ; dimension after PCA ($k$); plaintext space $\mathds{Z}_{2^l}$}
        	\Output{Dimension reduced matrix $\boldsymbol{X'}$}
            
            Data providers compute features' mean $\mathbf{S_{i}}$ locally, generate $M$ shares $\{ {\langle \mathbf{S_{i}} \rangle}_1, \cdots, {\langle \mathbf{S_{i}} \rangle}_M \} $, and distribute them;\\
            
            Data providers receive $\{ {\langle \mathbf{S_{1}} \rangle}_i, \cdots, {\langle \mathbf{S_{M}} \rangle}_i \}$ from each other, compute ${\langle \mathbf{V_{i}} \rangle} = \sum_{j=1}^M {\langle \mathbf{S_{j}} \rangle}_i$ mod $2^l$, and send it to server ;\\
            
            Sever receives $\{ {\langle \mathbf{V_{1}} \rangle}, \cdots, {\langle \mathbf{V_{M}} \rangle} \}$, computes $\overline{x} = \sum_{i=1}^M {\langle \mathbf{V_{i}} \rangle}$ mod $2^l$, and sends it to all data providers;\\
            
            Data providers receive $\overline{x}$ and normalize data columns to zero locally;\\
            
            Data providers compute $ \boldsymbol{C_i}  =  \frac{1}{n-1}\boldsymbol{X_i}^T\boldsymbol{X_i} $, generate $M$ shares $\{ {\langle \mathbf{C_{i}} \rangle}_1, \cdots, {\langle \mathbf{C_{i}} \rangle}_M \} $, and distribute them;\\
            
            Data providers receive $\{ {\langle \mathbf{C_{1}} \rangle}_i, \cdots, {\langle \mathbf{C_{M}} \rangle}_i \}$ from each other, compute ${\langle \mathbf{V'_i} \rangle} = \sum_{j=1}^M {\langle \mathbf{C_{j}} \rangle}_i$ mod $2^l$ locally, and send it to sever;\\
            
            Server receives $\{ {\langle \mathbf{V'_{1}} \rangle}, \cdots, {\langle \mathbf{V'_{M}} \rangle} \}$ and computes global covariance matrix ${ \mathbf{C}} = \sum_{i=1}^M {\langle \mathbf{V'_{i}} \rangle}$ mod $2^l$;\\
            
            Server computes eigenvectors, chooses $k$ of them to form the matrix $\boldsymbol{T}$, and sends it to data providers;\\
            
            Data providers receive matrix $\boldsymbol{T}$, calculate $\boldsymbol{X'_i} = \boldsymbol{X_iT}$ locally, and send it to data consumer;\\ 
            
            Data consumer receives dimension reduced matrix $\boldsymbol{X'}={(\boldsymbol{{X'_{1}}^T}, \cdots, \boldsymbol{{X'_{M}}}^T)}^T$ for modeling;\\
            
        	\Return{dimension reduced matrix $\boldsymbol{X'}$};
        	\caption{SS based PPPCA framework}
        	\label{alg: ss-part-p}
        \end{algorithm}

    \subsection{Security Discussion}
        Our protocols are secure against honest-but-curious adversaries. That is, data providers, server, and data consumer will strictly follow the protocol, but they will keep all intermediate computation results and try to infer as much information as possible. We also assume that the server does not collude with any data providers.
        
        In PPPCA, data providers only receive the mean of each feature which does not contain the details of each feature. 
        As for data provider $p$ in HE based method, all it receives from others are encrypted matrices.
        The server only has the covariance matrix which does not contain private information neither. 
        The data consumer gets the result of PPPCA, which can not infer private raw data. 
        Therefore, the security of our proposed PPPCA framework is satisfied.

\section{Experimental Studies}

    In this section, we focus on answering the following research questions. Q1: is the accuracy of the model built upon PPPCA the same as the model with PCA? Q2: what is the difference of efficiency between homomorphic encryption based method and secret sharing based method?

    
    \subsection{Experimental Setup}
      
        We conduct experiments on 4 datasets, including one real-world fraud detection dataset (Fraud) \cite{dataset} and 3 public datasets from UCI, i.e., APS Failure dataset (APS) \cite{dataset-aps}, Wine-quality dataset (Wine) \cite{CorCer09}, and Online News Popularity dataset (Online)  \cite{dataset-online-news}. The first 2 datasets are for classification problems and the others are for regression tasks. The fraud detection dataset has 298 features and 792,004 transactions. We focus on horizontally partitioned case, therefore we assume these data are hold by different parties and each of them has roughly equal partial samples. 
        
        We use five-fold cross validation during experiments. 
        After PPPCA, we adopt the most popular three kinds of models, i.e., linear model, tree based ensemble model, and deep model for classification and regression problem. Note that it can be generalized to any other machine learning models. We adopt Area Under the receiver operating characteristic Curve (AUC) as the evaluation metric for binary classification tasks and Root Mean Squared Error (RMSE) for regression tasks.
    
    \subsection{Comparison Results}
        \nosection{Answer to Q1: accuracy} We assume there are two parties and conduct the following comparison to answer Q1. 
        
        First, we compare the model performance built on (1) PCA using centralized plaintext data, (2) PCA using decentralized plaintext data, and (3) PPPCA using Homomorphic Encryption (HE) and Secret Sharing (SS). 
        We choose logistic regression model for fraud and APS, since they are binary classification tasks, and choose linear regression model for Wine and Online dataset, since they are regression tasks. 
        We report the comparison results in Table 1. 
        From it, we can see that the model 
        built upon PPPCA is lossless comparing with the model build on centralized plaintext PCA, and models build on local PCA separately have worse performance. This experiment indicates the effectiveness of PPPCA. 
            
        Second, we compare the performance of different models, i.e., Logistic Regression (LR), Gradient Boosting Decision Tree (GBDT), and Deep Neural Networks (DNN) on APS dataset. We summarize the results in Table 2, where we find similar results as in Table 1. This experiment indicates the PPPCA can be used to build any privacy preserving machine learning models for multi-parites. 
            
            \begin{table}[t]\label{re1}
                \footnotesize
                \caption{AUC and RMSE results of logistic regression and linear regression on different datasets}
                
                \begin{tabular}{|c|c|c|cc|}

                  \hline
                  Dataset & \begin{tabular}[c]{@{}c@{}}PCA\\ (Centralized)\end{tabular}  & \begin{tabular}[c]{@{}c@{}}PCA\\ (Separately)\end{tabular} & \multicolumn{2}{c|}{PPPCA} \\
                  \hline
                  Secure Op & -  & - & HE & SS   \\
                  \hline
                  Fraud & 0.9663  & 0.8978 & 0.9692 & 0.9613 \\
                  \hline
                  APS & 0.9915  & 0.9863 & 0.9918 & 0.9906    \\
                  \hline
                  Wine& 0.7122  & 0.7498  & 0.7103 & 0.7126   \\
                  \hline
                  Online & 8.0484 & 8.0801 & 8.0517 & 8.0493   \\
                  \hline

                \end{tabular}
            \end{table}




            \begin{table}[t]\label{re2}
                \footnotesize
            	\caption{AUC results of different models on APS dataset}
                \begin{tabular}{|c|c|c|cc|}

                  \hline
                  Model & \begin{tabular}[c]{@{}c@{}}PCA\\ (Centralized)\end{tabular} &  \begin{tabular}[c]{@{}c@{}}PCA\\ (Separately)\end{tabular}  & \multicolumn{2}{c|}{PPPCA} \\
                  \hline
                  Secure Op & -  & - & HE & SS   \\
                  \hline
                  LR & 0.9915  & 0.9863 & 0.9918 & 0.9906 \\
                  \hline
                  GBDT & 0.9941  & 0.9857  & 0.9943 & 0.9948  \\
                  \hline
                  DNN & 0.9937 & 0.9913  & 0.9928 & 0.9931  \\
                  \hline

                \end{tabular}
            \end{table}

        \nosection{Answer to Q2: efficiency}
        We now study the time efficiency of our two proposed methods. 
        To do this, we vary the number of parties and compare the running time of PPPCA on Wine dataset. 
        We report the results (in seconds) in Table \ref{tab:results}. 
        We can see that with the increase of data providers, the running time of HE based method grows slower than SS based method. This is because the time of HE is mainly spent on encryption, which can be done by each party in parallel. In contrast, as the number of parties increases, the running time of SS based method  increases dramatically.

            \begin{table}[t]
            	\centering
            	\caption{Running time (in seconds) of PPPCA with different number of parties}
            	\label{tab:results}
            	\small
            	 \begin{tabular}{|c|c|c|c|}
                    \hline
                    \diagbox{Method}{Runing time}{Num of parties}    &  2  &  3 &  4   \\ \hline
                    HE based method    &  3.16  &  3.23 &  3.36   \\ \hline
                    SS based method   &  2.91  &  3.22  &  3.42   \\ \hline
        
                \end{tabular}
            	
            \end{table}
  
\section{Conclusion}
    In this paper, we presented a general modeling paradigm with Privacy Preserving Principal Component Analysis (PPPCA) for horizontally partitioned data under server-aided setting. It involves two methods using homomorphic encryption and secret sharing respectively. Experiments
    on three UCI datasets and a real-world dataset demonstrated the efficiency and effectiveness of PPPCA. In the future, we would like to deploy PPPCA in real-world applications in Ant Financial.
    
\section{Acknowledgments}
    We would like to thank all the anonymous reviewers for their valuable suggestions. Y. Liu and G. Liu were supported in part by the Anhui Sun Create Electronics Company Ltd., under Grant KD1809300321, the National Key R\&D Program of China under Grant 2018YFC0832101, 
    and STCSM18\textit{DZ}2270700.



\bibliographystyle{IEEEtran}

\end{document}